\documentclass[a4paper,11pt,english]{article}

\usepackage{graphicx,amssymb,amstext,amsmath}
\usepackage[square,numbers,sort&compress]{natbib}
\usepackage{flafter}
\usepackage{enumerate}
\usepackage{courier}
\usepackage{subfigure}
\usepackage[draft]{fixme}
\usepackage{authblk}
\usepackage{hyperref} % load last

\newif\iffinal 
\usepackage{tikz}
\usetikzlibrary{arrows,shapes,snakes,automata,backgrounds,petri}
\usetikzlibrary{positioning}
\pgfrealjobname{note} 

\iffinal

\else

\fi
\newtheorem{lemma}{Lemma}
\newtheorem{theorem}{Theorem}
\newtheorem{proposition}{Proposition}

\newcommand\bigzero{\makebox(0,0){\text{\huge0}}}
\newcommand{\keywords}[1]{\smallskip\par\noindent {\small{\em Keywords\/}: #1}}

\begin{document}
\author[1,2]{Andreas Sand}
\affil[1]{Bioinformatics Research Centre, Aarhus University, Denmark}
\affil[2]{Department of Computer Science, Aarhus University, Denmark}
\author[3]{Mike Steel}
\affil[3]{Allan Wilson Centre for Molecular Ecology and Evolution, University of Canterbury, Christchurch, New Zealand}

\title{The standard lateral gene transfer model is statistically
  consistent for pectinate four-taxon trees}

\maketitle

\begin{abstract}
  Evolutionary events such as incomplete lineage sorting and lateral
  gene transfer constitute major problems for inferring species trees
  from gene trees, as they can sometimes lead to gene trees which
  conflict with the underlying species tree. One particularly simple
  and efficient way to infer species trees from gene trees under such
  conditions is to combine three-taxon analyses for several genes
  using a majority vote approach. For incomplete lineage sorting this
  method is known to be statistically consistent, however, in the case
  of lateral gene transfer it is known that a zone of inconsistency
  does exist for a specific four-taxon tree topology. In this paper we
  analyze all remaining four-taxon topologies and show that no other
  inconsistencies exist.

  \keywords{Phylogenetic trees, lateral gene transfer, statistical
    consistency}
\end{abstract}

\section{Introduction}

A major problem in inferring species trees from gene trees is that
different genes often suggest different evolutionary
histories~\cite{Galtier2008}. This phenomenon is caused by incomplete
lineage sorting and reticulate evolutionary events, i.e. hybridization
and lateral gene transfer, and it naturally poses the question,
whether the underlying gene tree can be consistently reconstructed
from a set of gene trees? In the case of hybridization, it is clear
that no single tree can adequately describe the evolution of the taxa
under study, and that a network is usually a more appropriate
representation. For incomplete lineage sorting recent theoretical work
based on the multi-species coalescent has shown that the most probable
gene tree topology can differ from the species tree topology, when the
number of taxa is greater than three~\cite{Sanderson2010}. By contrast
it has long been known that for triplets, the matching topology is the
most probable topology~\cite{Tajima1983, Nei1987}. Complementary to
this, it was recently proved that, under the standard or extended
models of lateral gene transfer (LGT), the matching gene tree topology
is also the most probable topology for a tree with three taxa; but
for the fork-shaped four-taxon tree topology there exist branch
lengths for which the matching topology of a triplet has the lowest
probability of the three possible topologies~\cite{Steel2013}. In this
paper we start by recalling the two models of lateral gene transfer
and the key definitions from~\cite{Steel2013}. We then give a thorough
analysis of the other four-taxon tree topology (the pectinate
topology), showing that in this case, the matching topology for a set
of three leaves is always the most probable topology, regardless of
the location of the fourth taxon. This completes the four-taxon case
and implies that four-taxon species trees can be consistently
reconstructed using a triplet-based majority vote approach, provided
that the branch lengths meet the conditions given in~\cite{Steel2013}.
\section{Key definitions}

Throughout this paper $X$ will denote a set of $n$ taxa, and $A$ will
be a subset of size $3$ of $X$. Let $T$ be a rooted phylogenetic
species tree with leaf set $X$ and root $\rho$. Regarding $T$ as a
1-dimensional simplicial complex so each point $p$ in $T$ is either a
vertex or an element of an interval that corresponds to an edge, we
use a coalescence time scale: $t : T \rightarrow [0, \infty)$ with
coalescence time increasing into the past, such that
\begin{itemize}
  \item{$t(p) = 0 \Leftrightarrow p\text{ is a leaf}$, and}
  \item{if $u$ is a descendant of $v$ then $t(u) < t(v)$.}
\end{itemize}

We will denote the time from the present to the most recent common
ancestor (MRCA) of the two most closely related taxa in $A$ by
$t_A$; e.g. if $T|A = a|bc$ then $t_A$ is the time from the present to
the MRCA of $b$ and $c$.

Linz et al. defines what we will refer to as the standard LGT model
in~\cite{Linz2007}. This model makes the following assumptions:
\begin{enumerate}
\item{A binary, labeled, rooted and clocklike species tree $T$ is
    given, as well as all the splitting times along this tree;}
\item{differences between a specific gene tree and $T$ are only caused
    by LGT events;}
\item{the transfer rate is homogeneous per gene and unit time;}
\item{genes are transfered independently;}
\item{one copy of the transferred gene still remains in the donor
    genome; and}
\item{the transferred gene replaces any existing orthologous
    counterpart in the acceptor genome.}
\end{enumerate}
Based on this model, the authors in~\cite{Steel2013} considered an
extended LGT model, in which the rate of gene transfer between two
lineages can be decreasing in the distance between the two
lineages. Specifically, letting $d(p, p')$ be the evolutionary
distance between contemporaneous points $p$ and $p'$ in $T$, item
three above is replaced by the following assumption:
\begin{enumerate}
\setcounter{enumi}{2}
\item{transfer events on $T$ occur as a Poisson process through time,
    in which the rate of transfer events from point $p$ on a lineage to
    a contemporaneous point $p'$ on another lineage at time $t$ occurs
    at rate $f(d(p, p'), t)$, where $f(d, t)$ is a monotone
    non-increasing function in $d$ (but can vary non-monotonically in
    $t$).}
\end{enumerate}

\subsection{Lateral gene transfer events and transfer sequences}

A \emph{lateral gene transfer (LGT) on $T$} is an arc from $p \in T$
to $p' \in T$ where $t(p) = t(p')$ and neither $p$ or $p'$ are
vertices of $T$. We write $\sigma = (p, p')$ to denote this transfer
event and we write $t(\sigma)$ for the common value of $t(p)$ and
$t(p')$. We will assume that no two transfer events occur at exactly
the same time.

Let $\underline{\sigma} = \sigma_1 \ldots \sigma_k$ be a sequence of
transfer events arranged in increasing $t$-value:
\begin{equation*}
  0 < t(\sigma_1) < t(\sigma_2) < \dots < t(\sigma_k) < t(\rho).
\end{equation*}

Given a species tree $T$ and a transfer sequence $\underline{\sigma} =
\sigma_1 \ldots \sigma_k$ on $T$, we obtain an associated \emph{gene
  tree} $T[\underline{\sigma}]$.  An LGT arc $\sigma$ from point $p$
to $p'$ in $T$ describes the event that the gene which was present on
the edge at $p'$ is replaced by the transfered gene from $p$. Thus, if
we trace the history of a gene from the present to the past, each time
we encounter an incoming horizontal arc into this edge, we follow this
arc (against the direction of the arc). Mathematically this is
formalized as follows: For a transfer sequence $\underline{\sigma} =
\sigma_1 \ldots \sigma_k$ where $\sigma_i = (p_i, p_i')$ consider the
tree $T$ together with a directed edge for each $\sigma_i$ placed
between $p_i$ and $p_i'$ for each $i \in\{ 1, \ldots, k \}$ and regard
this network as a one-dimensional simplicial complex. Now for each $i
\in \{1, \ldots, k \}$ delete the interval above $p_i'$ and consider
the minimal connected subgraph of the resulting complex that contains
$X$. This is $T[\underline{\sigma}]$.

Given the pair $T$, $\underline{\sigma} = \sigma_1 \ldots \sigma_k$,
define the following sequence of $X$--trees:
\begin{equation*}
  T_0 = T, T_r = T_{r-1}[\sigma_r].
\end{equation*}
And given $T' \in \{ T_0, T_1, \ldots, T_k \}$, a point $p \in T'$ and
a non-empty subset $Y$ of $X$, let $des_Y(T', p)$ denote the subset of
$Y$ whose elements are descendants of $p$ in $T'$.

\subsection{Triplet analysis}
Let $A$ be a subset of $X$ of size $3$, let $T$ be a phylogenetic
species tree on $X$, let $\underline{\sigma}$ be a sequence of
transfer events on $T$, and let $\sigma_r=(p_r,p_r')$ be a specific transfer
event on $T$. We say that:
\begin{itemize}
\item{$\underline{\sigma}$ induces a \emph{match} for $A$ if $T|A =
    T[\underline{\sigma}]|A$. Otherwise we say that $\underline{\sigma}$
    induces a \emph{mismatch} for $A$.}
\item{$\sigma_r$ is \emph{into an $A$-lineage} if $des_A(T, p_r')$ is a
  single element in $A$.}
\item{$\sigma_r$ is an \emph{$A$-transfer} and it transfers $x$
    if $des_A(T_{r-1}, p_r') = \{ x \}$ for some $x \in A$.}
\item{$\sigma_r$ is an \emph{$A$-moving transfer} and it moves
    $x$ if it transfers $x$ and $des_A(T_{r-1}, p_r) = \emptyset$.}
\item{$\sigma_r$ is an \emph{$A$-joining transfer} and it joins $x$ to
    $y$ if it transfers $x$ and $des_A(T_{r-1}, p_r) = \{ y \}$ for
    some $y \in A$.}
\end{itemize}
Note that any $A$--transfer is either an $A$-moving or an
$A$-joining transfer.

Let $\underline{\sigma} = \sigma_1, \sigma_2, \ldots, \sigma_k$ be a
sequence of transfer events on $T$ with $t(\sigma_k) < t_A$ and no
$A$-joining transfers. Then construct the sequence $T'_0, T'_1,
\ldots, T'_k$ of trees by the following procedure: Set $T'_0 = T$ and
construct $T'_{i+1}$ from $T'_i$:

\begin{itemize}
\item{
    If $\sigma_i$ is not $A$--moving
    \begin{enumerate}
    \item{$T'_{i} = T'_{i-1}$}
    \end{enumerate}
  }
\item{
    else if $\sigma_i = (p_i, p'_i)$ moves $x \in A$, let $T'_{i}$ be
    the tree obtained from $T_{i-1}'$ by
    \begin{enumerate}
      \item{deleting all $p \in T'_{i-1}$ with $t(p) < t(\sigma_i)$,}
      \item{labeling $p_i$ by $x$,}
      \item{for both $z \in A - \{ x \}$, assigning label $z$ to the
          unique point $p_z$ of $T'_{i-1}$ that has $t(p_z) = t(\sigma_i)$
          and $z \in des_A(T'_{i-1}, p_z)$, and}
      \item{regarding all other leaves in the tree as unlabeled.}
    \end{enumerate}
  }
  \end{itemize}

\noindent The following two lemmas were given and proved in~\cite{Steel2013}:
\begin{lemma}
  Let $\underline{\sigma}$ be a sequence of transfer events on a rooted
  binary $X$--tree $T$ and let $A = \{ a,b,c \} \subseteq X$.
  \begin{enumerate}
  \item{If $\underline{\sigma}$ induces a mismatch for $A$, then
      $\underline{\sigma}$ must contain an $A$--transfer with a
      $t$--value less that $t_A$.}
  \item{Moreover precisely one of the following occurs:
      \begin{enumerate}
      \item{$\underline{\sigma}$ has no $A$--transfers. In this
          case, $\underline{\sigma}$ induces a match for $A$.
          \label{lemma:no_A_transfers}}
      \item{$\underline{\sigma}$ contains at least one $A$--joining
          transfer. In this case, if the first such transfer in
          $\underline{\sigma}$ joins $x$ to $y$, then
          $T[\underline{\sigma}]|A = z|xy$ where $\{ x,y,x \} =
          A$.
          \label{lemma:at_least_on_A_joining}}
      \item{$\underline{\sigma}$ has no $A$--joining transfers, but it has an
          $A$--moving transfer with a $t$--value less than
          $t_A$. In this case, if $\sigma_r$ denotes the first
          such $A$--moving transfer in $\underline{\sigma}$ then:
          \begin{equation*}
            T[\underline{\sigma}]|A = T[\sigma_r, \ldots, \sigma_k]|A.
          \end{equation*}
          \label{lemma:no_A_joining}
        }
      \end{enumerate}
    }
  \end{enumerate}
  \label{lemma:match_mismatch}
\end{lemma}

\begin{lemma}
  Suppose $\underline{\sigma} = \sigma_1, \sigma_2, \ldots, \sigma_k$ is
  a sequence of transfer events on a rooted binary $X$--tree $T$ with
  $t(\sigma_k) < t_A$ and with no $A$--joining transfers. Then
  $T[\underline{\sigma}]|A = T'_k|A$.
  \label{lemma:T_prime_topology}
\end{lemma}

\section{Three-taxon trees}
For completeness we restate the following result for three-taxon trees
from~\cite{Steel2013}:
\begin{proposition}
  If T has just three taxa, then under the extended LGT model, the
  probability that a transfer sequence induces a match for the three
  taxa is strictly greater than the probability it induces either one
  of the two mismatch topologies (which have equal probability).
\end{proposition}

\section{Four-taxon trees}

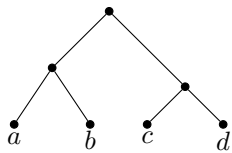
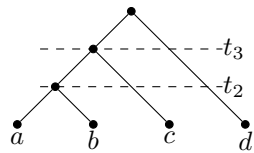
\begin{figure}[b]
  \centering
  \subfigure[The fork-shaped four-taxon tree topology.]{
    \makebox[.4\textwidth]{
  \beginpgfgraphicnamed{figures/fork_tree.tikz-external}%
  \begin{tikzpicture}[scale=0.5]
  \tikzstyle{every node} = [font=\small, circle, fill, draw, minimum size=1mm, inner sep=0pt]

  \draw (0,0) node (abcd) {};
  
  \draw (-1.5,-1.5) node (ab) {};
  \draw (2,-2) node (cd) {};

  \draw (-2.5,-3) node (a)[label=below:$a$] {};
  \draw (-0.5,-3) node (b)[label=below:$b$] {};
  \draw (1,-3) node (c)[label=below:$c$] {};
  \draw (3,-3) node (d)[label=below:$d$] {};

  \draw (abcd) -- (ab);
  \draw (abcd) -- (cd);
  \draw (ab) -- (a);
  \draw (ab) -- (b);
  \draw (cd) -- (c);
  \draw (cd) -- (d);

\end{tikzpicture}%
  \endpgfgraphicnamed%

    }
    \label{fig:fork_tree}
  }
  \hspace{1cm}
  \subfigure[The pectinate four-taxon tree topology $(ab;c;d)$]{
    \makebox[.4\textwidth]{
  \beginpgfgraphicnamed{figures/pectinate_tree.tikz-external}%
  \begin{tikzpicture}[scale=0.5]
  \tikzstyle{every node} = [font=\small, circle, fill, draw, minimum size=1mm, inner sep=0pt]
  
  \draw (0,0) node (abcd) {};
  
  \draw (-1,-1) node (abc) {};
  
  \draw (-2,-2) node (ab) {};

  \draw (-3,-3) node (a)[label=below:$a$] {};
  \draw (-1,-3) node (b)[label=below:$b$] {};
  \draw (1,-3) node (c)[label=below:$c$] {};
  \draw (3,-3) node (d)[label=below:$d$] {};

  \draw (abcd) -- (abc);
  \draw (abcd) -- (d);
  \draw (abc) -- (ab);
  \draw (abc) -- (c);
  \draw (ab) -- (a);
  \draw (ab) -- (b);

  \draw[draw=white, fill=white] (2.7, -1) node {$t_3$};
  \draw[dashed] (-2.4, -1) -- (2.4, -1);

  \draw[draw=white, fill=white] (2.7, -2) node {$t_2$};
  \draw[dashed] (-2.4, -2) -- (2.4, -2);

\end{tikzpicture}%
  \endpgfgraphicnamed%

    }
    \label{fig:pectinate_tree}
  }
  \caption{The two four-taxon tree topologies.}
  \label{fig:four_taxon_topologies}
\end{figure}

For four-taxon trees there are two rooted binary tree topologies --
the \emph{fork-shaped} topology with two cherries as shown in
Fig.~\ref{fig:fork_tree} and the \emph{pectinate} tree topology shown
in Fig.~\ref{fig:pectinate_tree}. The fork-shaped topology was studied
thoroughly in~\cite{Steel2013}, and we will study the pectinate tree
topology.

For four-taxon trees $a,b,c,d$, we will write $(ab; c; d)$ to denote
the pectinate tree topology depicted in
Fig.~\ref{fig:pectinate_tree}. This topology is symmetric to
$(ba;c;d)$, $(d;c;ab)$ and $(d;c;ba)$, but no other symmetries
hold. For any pectinate four-taxon tree we denote the time of the MRCA
of the two most closely related taxa by $t_{2}$, and the time of the
MRCA of the three most closely related taxa by $t_3$. Thus, for
example if the tree has topology $(ab;c;d)$, $t_2$ is the time of the
MRCA of $a$ and $b$, and $t_3$ is the time of the MRCA of $a$, $b$ and
$c$.

The main result in this paper is the following theorem:

\begin{theorem}
  Suppose $T$ is a pectinate four-taxon tree and $A = \{ a,b,c \}$ is
  a subset of the leaf set $X$ of $T$, and suppose that $T|A =
  ab|c$. Let $\underline{\sigma} = \sigma_1, \sigma_2, \ldots,
  \sigma_k$ be a random sequence of transfer events on $T$ generated
  by the standard LGT model of~\cite{Linz2007}, in which the rate of
  transfer events from point $p$ to a contemporaneous point $p'$ is
  $\lambda$. Then the probability that $\underline{\sigma}$ induces a
  match on $A$ is strictly higher than the probability that it induces
  either one of the two mismatch topologies (which have equal
  probability).

  More specifically: Let $\xi_{ab|c}$, $\xi_{ac|b}$ and $\xi_{bc|a}$
  denote the disjoint events that $\underline{\sigma}$ induces a tree
  displaying the triplet topologies $ab|c$, $ac|b$ or $bc|a$,
  respectively, and let $\mathbb{P}(\xi_{ab|c})$, $\mathbb{P}(\xi_{ac|b})$ and
  $\mathbb{P}(\xi_{bc|a})$ be the probabilities of these. Then for $\mu =
  \frac{1}{3} \lambda t_2$ and $B = 3 \lambda (t_3 - t_2)$ 
  \begin{enumerate}[(i)]
  \item{if $T$ is of type $(ab;c;*)$ we have:
      \begin{equation}
        \mathbb{P}(\xi_{ab|c}) = \frac{1}{3}(1 + e^{-7\mu} (\frac{3}{4} e^{-B} e^{-4\mu} + (1 - \frac{1}{2} e^{-B}) e^{-2\mu} + (1 - \frac{1}{4} e^{-B}))),
      \end{equation}
      and $\mathbb{P}(\xi_{ab|c}) > \frac{1}{3} > \mathbb{P}(\xi_{ac|b}) = \mathbb{P}(\xi_{bc|a})$
      for all values of $t_2$ and $t_3$;}
  \item{if $T$ is of type $(ab;*;c)$ we have:
      \begin{equation}
        \mathbb{P}(\xi_{ab|c}) = \frac{1}{3} ( 1 - e^{-7\mu} (\frac{3}{4}e^{-B} e^{-4\mu} - (1 - \frac{1}{2}e^{-B})e^{-2\mu} - (1 + \frac{5}{4}e^{-B})  ) ),
      \end{equation}
      and $\mathbb{P}(\xi_{ab|c}) > \frac{1}{3} > \mathbb{P}(\xi_{ac|b}) = \mathbb{P}(\xi_{bc|a})$
      for all values of $t_2$ and $t_3$; and}
  \item{if $T$ is of type $(a*;b;c)$ or $(b*;a;c)$ we have:
      \begin{equation}
        \mathbb{P}(\xi_{ab|c}) = \frac{1}{3} ( 1 + e^{-7\mu} (\frac{3}{8}e^{-B} e^{-4\mu} - (\frac{1}{2} - \frac{1}{4}e^{-B})e^{-2\mu} + (\frac{1}{2} + \frac{11}{8}e^{-B})  ) ),
      \end{equation}
      and $\mathbb{P}(\xi_{ab|c}) > \frac{1}{3} > \mathbb{P}(\xi_{ac|b}) = \mathbb{P}(\xi_{bc|a})$
      for all values of $t_2$ and $t_3$.}
  \end{enumerate}
  \label{theorem}
\end{theorem}

The proof of Theorem~\ref{theorem} relies on the analysis of the
discrete $7$-state Markov chain whose transition digraph is
illustrated in Fig.~\ref{fig:7StateModel}. We will therefore study
this Markov chain thoroughly, before we dive into the proof of
the theorem.

\subsection{The 7-state continuous-time Markov chain}
\label{sec:7-state}
\begin{figure}[t]
  \centering
  \begin{tikzpicture}[node distance=1.7cm,>=stealth',bend angle=45,auto,scale=0.7]
  \tikzstyle{every node} = [swap]
  \tikzstyle{state} = [font=\small, circle, draw, minimum size=7mm, inner sep=0pt]
  
  \node (1)[state] {$0$};
  \node (2)[state, below of=1] {$1$};
  \node (3)[state, below of=2] {$2$};
  \node (4)[state, below of=3] {$3$};
  \node (5)[state, below of=4] {$4$};
  \node (6)[state, below of=5] {$5$};
  \node (7)[state, below of=6] {$6$};

  \path[->] (1.south west) edge[bend right] node {$2$} (2.north west);
  \path[->] (2.south west) edge[bend right] node {$1$} (3.north west);
  \path[->] (3.south west) edge[bend right] node {$1$} (4.north west);
  \path[->] (4.south west) edge[bend right] node {$1$} (5.north west);
  \path[->] (5.south west) edge[bend right] node {$1$} (6.north west);
  \path[->] (6.south west) edge[bend right] node {$1$} (7.north west);

  \path[->] (7.north east) edge[bend right] node {$2$} (6.south east);
  \path[->] (6.north east) edge[bend right] node {$1$} (5.south east);
  \path[->] (5.north east) edge[bend right] node {$1$} (4.south east);
  \path[->] (4.north east) edge[bend right] node {$1$} (3.south east);
  \path[->] (3.north east) edge[bend right] node {$1$} (2.south east);
  \path[->] (2.north east) edge[bend right] node {$1$} (1.south east);

  \path[->] (1.west) edge[bend right] node {$1$} (7.west);

  \path[->] (3.west) edge[bend right] node {$1$} (5.west);

  \path[->] (5.east) edge[bend right] node {$1$} (3.east);

  \path[->] (7.east) edge[bend right] node {$1$} (1.east);
  
\end{tikzpicture}
  \caption{The 7 state continuous-time Markov model.}
  \label{fig:7StateModel}
\end{figure}
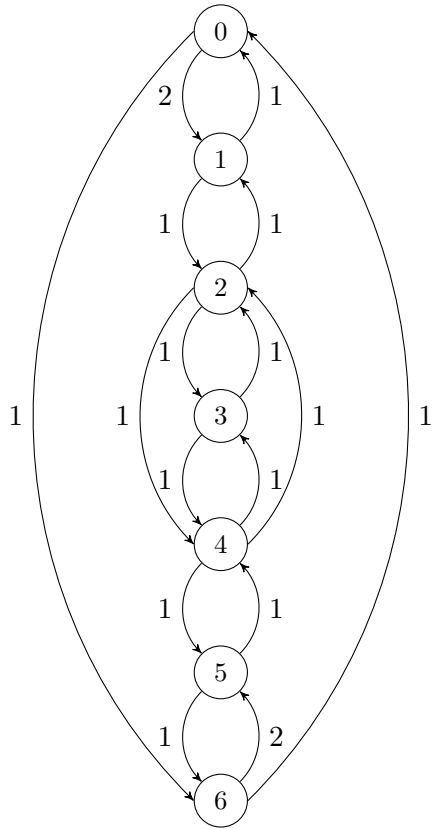

Let $Z_t : t \ge 0$ be the $7$--state continuous-time Markov chain
defined by the rate matrix
\begin{equation*}
  Q = 
  \left[ 
    \begin{array}{ccccccc}
     -3& 2 & 0 & 0 & 0 & 0 & 1 \\
      1 &-2& 1 & 0 & 0 & 0 & 0 \\
      0 & 1 &-3& 1 & 1 & 0 & 0 \\
      0 & 0 & 1 &-2& 1 & 0 & 0 \\
      0 & 0 & 1 & 1 &-3& 1 & 0 \\
      0 & 0 & 0 & 0 & 1 &-2& 1 \\
      1 & 0 & 0 & 0 & 0 & 2 &-3 \\
    \end{array} 
  \right]
\end{equation*}
and illustrated in Fig.~\ref{fig:7StateModel}, let $p_r(t) = \mathbb{P}(Z_t = r)$,
and let $\mathbf{p}(t) = [p_0(t), \ldots , p_6(t)]$. Then by standard
Markov chain theory~\cite{Stewart1994}
\begin{equation*}
  \frac{d}{dt}\mathbf{p(t)} = \mathbf{p(t)}Q\quad\text{ and }\quad\mathbf{p}(t) = \mathbf{p}(0)\exp(Qt).
\end{equation*}

The Markov chain is easily seen to be irreducible and ergodic and thus
it has a stationary distribution $\boldsymbol{\pi}$. To find
$\boldsymbol{\pi} $ let
\begin{equation*}
  P =
  \left[ 
    \begin{array}{ccccccc}
      0               & \frac{2}{3} & 0               & 0               & 0               & 0               & \frac{1}{3} \\
      \frac{1}{2} & 0               & \frac{1}{2} & 0               & 0               & 0               & 0 \\
      0               & \frac{1}{3} & 0               & \frac{1}{3} & \frac{1}{3} & 0               & 0 \\
      0               & 0               & \frac{1}{2} & 0               & \frac{1}{2} & 0               & 0 \\
      0               & 0               & \frac{1}{3} & \frac{1}{3} & 0               & \frac{1}{3} & 0 \\
      0               & 0               & 0               & 0               & \frac{1}{2} & 0               & \frac{1}{3} \\
      \frac{1}{3} & 0               & 0               & 0               & 0               & \frac{2}{3} & 0 \\
    \end{array} 
  \right]
\end{equation*}
be the transition probability matrix of $Q$ defined by
\begin{equation*}
  p_{ij} = \left\{
      \begin{array}{cl}
        \frac{q_{ij}}{\sum_{k \neq i} q_{ik}} &\mbox{ if $i \neq j$}\\
        0                                               &\mbox{ otherwise}
      \end{array} \right.
\end{equation*}
Then $\boldsymbol{\pi}$ can be computed as 
\begin{equation*}
  \boldsymbol{\pi} = \frac{-\boldsymbol{\phi}~ D_Q^{-1}}{|| \boldsymbol{\phi} D_Q^{-1} ||_1},
\end{equation*}
where $\boldsymbol{\phi}$ is the unique solution summing to $1$ of
$\boldsymbol{\phi}(I - P) = \mathbf{0}$ and $D_Q$ is the $7 \times 7$
diagonal matrix having the same diagonal as
$Q$~\cite{Stewart1994}. Using this we find
\begin{equation*}
  \boldsymbol{\pi} = \left( \frac{1}{12}, \frac{2}{12}, \frac{2}{12}, \frac{2}{12}, \frac{2}{12}, \frac{2}{12}, \frac{1}{12} \right).
\end{equation*}
% Computing $\mathbf{p}(9)$ and $\mathbf{p}(10)$ with $\mathbf{p}(0) = (1, 0, 0, 0, 0, 0,
% 0)$, we get
% \begin{equation*}
%   \mathbf{p}(9) \simeq \mathbf{p}(10) \simeq \left( \frac{1}{12}, \frac{2}{12}, \frac{2}{12}, \frac{2}{12}, \frac{2}{12}, \frac{2}{12}, \frac{1}{12} \right)
% \end{equation*}
% (and similarly for larger $t$), confirming that $\lim_{t \rightarrow \infty} \mathbf{p}(t) = \boldsymbol{\pi}$.

\noindent The eigenvalues of $Q$ are
\begin{align*}
  \lambda_1 = -5, \quad \lambda_2 &= -4, \quad \lambda_3 = -4\\
  \lambda_4 = -3, \quad \lambda_5 &= -1, \quad \lambda_6 = -1\\
  \lambda_7 &= 0
\end{align*}
with corresponding eigenvectors:
\begin{align*}
  \mathbf{v_1} &= (-2,  1,-1,  0,  1,-1, 2)\\
  \mathbf{v_2} &= (-1,  0,  1,  0,-1,  0, 1)\\ 
  \mathbf{v_3} &= (-2,  1,  0,  1,-2,  1, 0)\\
  \mathbf{v_4} &= (  2,-1,-1,  2,-1,-1, 2)\\
  \mathbf{v_5} &= (  4,  3,-1,-3,-2,  0, 2)\\
  \mathbf{v_6} &= (-2,-2,  0,  1,  1,  1, 0)\\
  \mathbf{v_7} &= (  1,  1,  1,  1,  1,  1, 1)  
\end{align*}
Thus every element $p_r(t)$ for $r = 0, 1, \ldots, 6$ and $t \ge 0$ is of the form:
\begin{equation*}
  p_r(t) = a_r + b_r\exp(-t) + c_r\exp(-3t) + d_r\exp(-4t) + e_r\exp(-5t)
\end{equation*}
for constants $a_j, \dots, d_j$ depending on $\mathbf{p}(0)$. To find these
constants we will solve the set of linear equations given by
\begin{equation*}
 \mathbf{p}(t) = \left[ e^{-5t} , e^{-4t}, e^{-4t}, e^{-3t}, e^{-1t}, e^{-1t}, 1 \right] X,
\end{equation*}
where $X$ is a $7 \times 7$ matrix containing the constants.  Let $D$
be the $7 \times 7$ diagonal matrix with the eigenvalues of $Q$,
$\lambda_1, \lambda_2, \ldots, \lambda_7$, being the diagonal entries,
and let $V$ be the $7 \times 7$ matrix with the eigenvector $\mathbf{v_i}$
corresponding to $\lambda_i$ being column $i$ of $V$. Then $Q =
VDV^-1$, and we get
\begin{align*}
  \mathbf{p}(t) &= \mathbf{p}(0)\exp(Qt)\\
  &= \mathbf{p}(0)\exp(VDV^{-1}t)\\
  &= \mathbf{p}(0) V \exp(Dt)V^{-1}\\
  &= \mathbf{p}(0) V 
  \left[ 
    \begin{array}{ccccccc}
      e^{-5t} &               &           &           &               &           &   \\
                & e^{-4t}     &           &           & \bigzero &           &   \\
                &               & e^{-4t} &           &               &           &   \\
                &               &           & e^{-3t} &               &           &   \\
                &               &           &           & e^{-t}       &           &   \\
                & \bigzero&            &           &               & e^{-t}  &   \\
                &               &           &           &               &           & 1
    \end{array} 
  \right]
  V^{-1}.
\end{align*}
Thus the set of linear equations we need to solve reduces to
\begin{equation*}
  \left[ 
    \begin{array}{c}
      e^{-5t}\\
      e^{-4t}\\
      e^{-4t}\\
      e^{-3t}\\
      e^{-1t}\\
      e^{-1t}\\
      1 
    \end{array}
  \right]^T X = 
  \mathbf{p}(0) V 
  \left[ 
    \begin{array}{ccccccc}
      e^{-5t} &               &           &           &               &           &   \\
                & e^{-4t}     &           &           & \bigzero &           &   \\
                &               & e^{-4t} &           &               &           &   \\
                &               &           & e^{-3t} &               &           &   \\
                &               &           &           & e^{-1t}     &           &   \\
                & \bigzero&            &           &               & e^{-1t} &   \\
                &               &           &           &               &           & 1
    \end{array} 
  \right]
  V^{-1}.
\end{equation*}
Doing this with $\mathbf{p}(0) = (1,0,0,0,0,0,0)$, we get
\begin{equation}
  \mathbf{p}(t) = 
  \frac{1}{12}
  \left[
    \begin{array}{c}
      e^{-5t}\\
      e^{-4t}\\
      e^{-4t}\\
      e^{-3t}\\
      e^{-1t}\\
      e^{-1t}\\
      1 
    \end{array}
  \right]^T
  \left[
    \begin{array}{ccccccc}
      3 & -3 &  3 &  0 & -3 &  3 & -3\\
      1 &  2 & -6 &  2 &  2 &  2 & -3\\
      2 & -4 &  4 & -4 &  4 & -4 &  2\\
      2 & -2 & -2 &  4 & -2 & -2 &  2\\
      2 & -2 & -6 & -8 & -2 & 10 &  6\\
      1 &  7 &  5 &  4 & -1 &-11 & -5\\
      1 &  2 &  2 &  2 &  2 &  2 &  1
   \end{array}
  \right].
  \label{eq:matrix_representation1}
\end{equation}
From this representation of $\mathbf{p}(t)$ we immediately get the following
Lemma, which will be usefull in the process of proving
Theorem~\ref{theorem}.
\begin{lemma}
  For all $t > 0$ and $\mathbf{p}(0) = (1,0,0,0,0,0,0)$
  \begin{equation*}
    \begin{aligned}
      ~&p_1(t) = -\frac{1}{4}e^{-5t} - \frac{1}{6} e^{-4t} - \frac{1}{6} e^{-3t} + \frac{5}{12}e^{-t} +\frac{1}{6},\\
      ~&p_5(t) = \frac{1}{4} e^{-5t} - \frac{1}{6} e^{-4t} - \frac{1}{6} e^{-3t} - \frac{1}{12} e^{-t} + \frac{1}{6},\\
      ~&p_0(t) + p_6(t) = \frac{1}{6} e^{-4t} + \frac{1}{3}e^{-3t} + \frac{1}{3}e^{-t} + \frac{1}{6}\text{, and}\\
      ~&p_1(t) + p_3(t) + p_5(t) = \frac{1}{2} - \frac{1}{2} e^{-4t}.
    \end{aligned}
  \end{equation*}
  \label{lemma:matrix_representation1}
\end{lemma}
Similarly if $\mathbf{p}(0) = (0,0,0,0,0,1,0)$ we get
\begin{equation}
  \mathbf{p}(t) = 
  \frac{1}{24}
  \left[
    \begin{array}{c}
      e^{-5t}\\
      e^{-4t}\\
      e^{-4t}\\
      e^{-3t}\\
      e^{-1t}\\
      e^{-1t}\\
      1 
    \end{array}
  \right]^T
  \left[
    \begin{array}{ccccccc}
      3 & -3 &  3 &  0  & -3 &  3 & -3\\
      0 &  0 &   0 &  0  &   0 &  0 &   0\\
    -2 &  4 & -4 &  4  & -4 &  4 & -2\\
    -2 &  2 &   2 &-4  &   2 &  2 & -2\\
      0 &  0 &   0 &   0 &   0 &  0 &   0\\
    -1 &-7 & -5 & -4 &   1 & 11&   5\\
      2 &  4 &   4 &   4 &   4 &  4 &   2
   \end{array}
  \right],
  \label{eq:matrix_representation2}
\end{equation}
and the following Lemma follows immediately:
\begin{lemma}
  For all $t > 0$ and $\mathbf{p}(0) = (0,0,0,0,0,1,0)$
  \begin{align*}
  ~&p_5(t) = \frac{1}{8} e^{-5t} + \frac{1}{6} e^{-4t} + \frac{1}{12} e^{-3t} + \frac{11}{24} e^{-t} + \frac{1}{6},\\
  ~&p_0(t) + p_6(t) = -\frac{1}{6} e^{-4t} - \frac{1}{6}e^{-3t} + \frac{1}{6}e^{-t} + \frac{1}{6}\text{, and}\\
  ~&p_1(t) + p_3(t) + p_5(t) = \frac{1}{2} e^{-4t} + \frac{1}{2}.
  \end{align*}
  \label{lemma:matrix_representation2}
\end{lemma}
\subsection{Proof of part \textit{(i)}}
\label{sec:proof_i}

Let $T$ be a four-taxon tree over the set of taxa $X = \{ a, b, c,
*\}$ with topology $(ab;c;*)$ (here $*$ refers to the fourth taxon,
the identity of which plays no role when we come to consider the
topology of the triple $a, b, c$), let $A = \{ a, b, c \}$, and let
$\underline{\sigma} = \sigma_1, \sigma_2, \ldots, \sigma_k$ be a
random sequence of transfer events on $T$ generated by the standard
LGT model in which the rate of transfer events from point $p$ to point
$p'$ is $\lambda$.  Let $\xi$ be any one of the three events
$\xi_{ab|c}$, $\xi_{ac|b}$ or $\xi_{bc|a}$, and let $J$ denote the
(stochastic) number of $A$--joining transfers between time $t = 0$ and
time $t = t_2$. Then by the law of total probability
\begin{equation}
    \begin{aligned}
      \mathbb{P}(\xi) &= \mathbb{P}(\xi,  J > 0) + \mathbb{P}(\xi, J = 0)\\
      &= \mathbb{P}(\xi | J > 0) \mathbb{P}(J > 0) + \mathbb{P}(\xi | J = 0) \mathbb{P}(J = 0).
    \end{aligned}
\end{equation}
To find $\mathbb{P}(J > 0)$ and $\mathbb{P}(J = 0)$ we observe that $J$ has a Poisson
distribution with mean $2 \lambda t_2$, since at any moment in the
interval $[0, t_2]$, there are four lineages, three of which lead
to leaves in $A$, and for each of these, the rate of transfer from
that $A$--lineage to another $A$--lineage is $\lambda \cdot 2/3$. This
means that the cumulative rate of an $A$--joining transfer is $3 \cdot
\lambda \cdot 2/3 = 2\lambda$ at any given time in the interval $[0,
t_2]$. Thus
\begin{equation*}
  \mathbb{P}(J = 0) = e^{-2\lambda t_2}\quad\text{and}\quad \mathbb{P}(J > 0) = 1 - e^{-2\lambda t_2},
\end{equation*}
and we arrive at
\begin{equation}
  \mathbb{P}(\xi) = \mathbb{P}(\xi | J > 0) (1 - e^{-2 \lambda t_2}) + \mathbb{P}(\xi | J = 0) e^{-2\lambda t_2},
  \label{eq:P_xi}
\end{equation}
where $\xi$ is any one of the events $\xi_{ab|c}$, $\xi_{ac|b}$ or
$\xi_{bc|a}$.  We will now consider the two factors $\mathbb{P}(\xi | J > 0)$ and
$\mathbb{P}(\xi | J = 0)$ in turn.

{$\boldsymbol{\mathbb{P}(\xi | J > 0)}$:}
Lemma~\ref{lemma:match_mismatch}.\ref{lemma:at_least_on_A_joining}
tells us that if there is at least one $A$--joining transfer in
$\underline{\sigma}$, then the first one of these decides the
resulting topology of $T[\underline{\sigma}]|A$. There are $6$
possibilities for this first $A$--joining transfer: $a \rightarrow b$,
$a \leftarrow b$, $a \rightarrow c$, $a \leftarrow c$, $b \rightarrow
c$ and $b \leftarrow c$. The first two of these will give
$T[\underline{\sigma}]|A = ab|c$, the next two will give
$T[\underline{\sigma}]|A = ac|b$, while the last two will give
$T[\underline{\sigma}]|A = bc|a$. As they are all equally likely, we
get
\begin{equation}
  \mathbb{P}(\xi | J > 0) = \frac{1}{3}.
  \label{eq:P_xi_J_gt_zero}
\end{equation}

\begin{figure}[h]
  \centering
  \beginpgfgraphicnamed{figures/12cycle.tikz-external}%
  \begin{tikzpicture}[scale=0.8]
\tikzstyle{every node} = [font=\small, circle, fill, draw, minimum size=1mm, inner sep=0pt]
\tikzset{>=stealth}
\begin{scope}[shift={(-0.750000,5.625000)}]
	\draw (0, 0.0) node(a)[label=below:$a$] {};
	\draw (0.5, 0) node(b)[label=below:$b$] {};
	\draw (1.0, 0) node(c)[label=below:$c$] {};
	\draw (1.5, 0) node(d)[label=below:$*$] {};
	\draw (0.25, 0.25) node(ab) {};
	\draw (0.50, 0.50) node(abc) {};
	\draw (0.75, 0.75) node(abcd) {};
	\draw (abcd) -- (abc);
	\draw (abcd) -- (d);
	\draw (abc) -- (ab);
	\draw (abc) -- (c);
	\draw (ab) -- (a);
	\draw (ab) -- (b);
\end{scope}
\draw[<->] (1.247470, 5.868886) -- (2.440420, 5.481273);
	\draw (0.000000, 7.250000) node[fill={none}, minimum size=5mm] {1};
\begin{scope}[shift={(2.250000,4.821152)}]
	\draw (0, 0.0) node(a)[label=below:$a$] {};
	\draw (0.5, 0) node(b)[label=below:$*$] {};
	\draw (1.0, 0) node(c)[label=below:$c$] {};
	\draw (1.5, 0) node(d)[label=below:$b$] {};
	\draw (0.25, 0.25) node(ab) {};
	\draw (0.50, 0.50) node(abc) {};
	\draw (0.75, 0.75) node(abcd) {};
	\draw (abcd) -- (abc);
	\draw (abcd) -- (d);
	\draw (abc) -- (ab);
	\draw (abc) -- (c);
	\draw (ab) -- (a);
	\draw (ab) -- (b);
\end{scope}
\draw[<->] (4.014784, 4.458869) -- (4.854102, 3.526712);
	\draw (3.750000, 6.245191) node[fill={none}, minimum size=5mm] {2};
\begin{scope}[shift={(4.446152,2.625000)}]
	\draw (0, 0.0) node(a)[label=below:$a$] {};
	\draw (0.5, 0) node(b)[label=below:$c$] {};
	\draw (1.0, 0) node(c)[label=below:$*$] {};
	\draw (1.5, 0) node(d)[label=below:$b$] {};
	\draw (0.25, 0.25) node(ab) {};
	\draw (0.50, 0.50) node(abc) {};
	\draw (0.75, 0.75) node(abcd) {};
	\draw (abcd) -- (abc);
	\draw (abcd) -- (d);
	\draw (abc) -- (ab);
	\draw (abc) -- (c);
	\draw (ab) -- (a);
	\draw (ab) -- (b);
\end{scope}
\draw[<->] (5.706339, 1.854102) -- (5.967131, 0.627171);
	\draw (6.495191, 3.500000) node[fill={none}, minimum size=5mm] {3};
\begin{scope}[shift={(5.250000,-0.375000)}]
	\draw (0, 0.0) node(a)[label=below:$*$] {};
	\draw (0.5, 0) node(b)[label=below:$c$] {};
	\draw (1.0, 0) node(c)[label=below:$a$] {};
	\draw (1.5, 0) node(d)[label=below:$b$] {};
	\draw (0.25, 0.25) node(ab) {};
	\draw (0.50, 0.50) node(abc) {};
	\draw (0.75, 0.75) node(abcd) {};
	\draw (abcd) -- (abc);
	\draw (abcd) -- (d);
	\draw (abc) -- (ab);
	\draw (abc) -- (c);
	\draw (ab) -- (a);
	\draw (ab) -- (b);
\end{scope}
\draw[<->] (5.868886, -1.247470) -- (5.481273, -2.440420);
	\draw (7.500000, -0.250000) node[fill={none}, minimum size=5mm] {4};
\begin{scope}[shift={(4.446152,-3.375000)}]
	\draw (0, 0.0) node(a)[label=below:$b$] {};
	\draw (0.5, 0) node(b)[label=below:$c$] {};
	\draw (1.0, 0) node(c)[label=below:$a$] {};
	\draw (1.5, 0) node(d)[label=below:$*$] {};
	\draw (0.25, 0.25) node(ab) {};
	\draw (0.50, 0.50) node(abc) {};
	\draw (0.75, 0.75) node(abcd) {};
	\draw (abcd) -- (abc);
	\draw (abcd) -- (d);
	\draw (abc) -- (ab);
	\draw (abc) -- (c);
	\draw (ab) -- (a);
	\draw (ab) -- (b);
\end{scope}
\draw[<->] (4.458869, -4.014784) -- (3.526712, -4.854102);
	\draw (6.495191, -4.000000) node[fill={none}, minimum size=5mm] {5};
\begin{scope}[shift={(2.250000,-5.571152)}]
	\draw (0, 0.0) node(a)[label=below:$b$] {};
	\draw (0.5, 0) node(b)[label=below:$*$] {};
	\draw (1.0, 0) node(c)[label=below:$a$] {};
	\draw (1.5, 0) node(d)[label=below:$c$] {};
	\draw (0.25, 0.25) node(ab) {};
	\draw (0.50, 0.50) node(abc) {};
	\draw (0.75, 0.75) node(abcd) {};
	\draw (abcd) -- (abc);
	\draw (abcd) -- (d);
	\draw (abc) -- (ab);
	\draw (abc) -- (c);
	\draw (ab) -- (a);
	\draw (ab) -- (b);
\end{scope}
\draw[<->] (1.854102, -5.706339) -- (0.627171, -5.967131);
	\draw (3.750000, -6.745191) node[fill={none}, minimum size=5mm] {6};
\begin{scope}[shift={(-0.750000,-6.375000)}]
	\draw (0, 0.0) node(a)[label=below:$a$] {};
	\draw (0.5, 0) node(b)[label=below:$b$] {};
	\draw (1.0, 0) node(c)[label=below:$*$] {};
	\draw (1.5, 0) node(d)[label=below:$c$] {};
	\draw (0.25, 0.25) node(ab) {};
	\draw (0.50, 0.50) node(abc) {};
	\draw (0.75, 0.75) node(abcd) {};
	\draw (abcd) -- (abc);
	\draw (abcd) -- (d);
	\draw (abc) -- (ab);
	\draw (abc) -- (c);
	\draw (ab) -- (a);
	\draw (ab) -- (b);
\end{scope}
\draw[<->] (-1.854102, -5.706339) -- (-0.627171, -5.967131);
	\draw (0.000000, -7.750000) node[fill={none}, minimum size=5mm] {7};
\begin{scope}[shift={(-3.750000,-5.571152)}]
	\draw (0, 0.0) node(a)[label=below:$a$] {};
	\draw (0.5, 0) node(b)[label=below:$*$] {};
	\draw (1.0, 0) node(c)[label=below:$b$] {};
	\draw (1.5, 0) node(d)[label=below:$c$] {};
	\draw (0.25, 0.25) node(ab) {};
	\draw (0.50, 0.50) node(abc) {};
	\draw (0.75, 0.75) node(abcd) {};
	\draw (abcd) -- (abc);
	\draw (abcd) -- (d);
	\draw (abc) -- (ab);
	\draw (abc) -- (c);
	\draw (ab) -- (a);
	\draw (ab) -- (b);
\end{scope}
\draw[<->] (-4.458869, -4.014784) -- (-3.526712, -4.854102);
	\draw (-3.750000, -6.745191) node[fill={none}, minimum size=5mm] {8};
\begin{scope}[shift={(-5.946152,-3.375000)}]
	\draw (0, 0.0) node(a)[label=below:$a$] {};
	\draw (0.5, 0) node(b)[label=below:$c$] {};
	\draw (1.0, 0) node(c)[label=below:$b$] {};
	\draw (1.5, 0) node(d)[label=below:$*$] {};
	\draw (0.25, 0.25) node(ab) {};
	\draw (0.50, 0.50) node(abc) {};
	\draw (0.75, 0.75) node(abcd) {};
	\draw (abcd) -- (abc);
	\draw (abcd) -- (d);
	\draw (abc) -- (ab);
	\draw (abc) -- (c);
	\draw (ab) -- (a);
	\draw (ab) -- (b);
\end{scope}
\draw[<->] (-5.868886, -1.247470) -- (-5.481273, -2.440420);
	\draw (-6.495191, -4.000000) node[fill={none}, minimum size=5mm] {9};
\begin{scope}[shift={(-6.750000,-0.375000)}]
	\draw (0, 0.0) node(a)[label=below:$c$] {};
	\draw (0.5, 0) node(b)[label=below:$*$] {};
	\draw (1.0, 0) node(c)[label=below:$b$] {};
	\draw (1.5, 0) node(d)[label=below:$a$] {};
	\draw (0.25, 0.25) node(ab) {};
	\draw (0.50, 0.50) node(abc) {};
	\draw (0.75, 0.75) node(abcd) {};
	\draw (abcd) -- (abc);
	\draw (abcd) -- (d);
	\draw (abc) -- (ab);
	\draw (abc) -- (c);
	\draw (ab) -- (a);
	\draw (ab) -- (b);
\end{scope}
\draw[<->] (-5.706339, 1.854102) -- (-5.967131, 0.627171);
	\draw (-7.500000, -0.250000) node[fill={none}, minimum size=5mm] {10};
\begin{scope}[shift={(-5.946152,2.625000)}]
	\draw (0, 0.0) node(a)[label=below:$b$] {};
	\draw (0.5, 0) node(b)[label=below:$c$] {};
	\draw (1.0, 0) node(c)[label=below:$*$] {};
	\draw (1.5, 0) node(d)[label=below:$a$] {};
	\draw (0.25, 0.25) node(ab) {};
	\draw (0.50, 0.50) node(abc) {};
	\draw (0.75, 0.75) node(abcd) {};
	\draw (abcd) -- (abc);
	\draw (abcd) -- (d);
	\draw (abc) -- (ab);
	\draw (abc) -- (c);
	\draw (ab) -- (a);
	\draw (ab) -- (b);
\end{scope}
\draw[<->] (-4.014784, 4.458869) -- (-4.854102, 3.526712);
	\draw (-6.495191, 3.500000) node[fill={none}, minimum size=5mm] {11};
\begin{scope}[shift={(-3.750000,4.821152)}]
	\draw (0, 0.0) node(a)[label=below:$b$] {};
	\draw (0.5, 0) node(b)[label=below:$*$] {};
	\draw (1.0, 0) node(c)[label=below:$c$] {};
	\draw (1.5, 0) node(d)[label=below:$a$] {};
	\draw (0.25, 0.25) node(ab) {};
	\draw (0.50, 0.50) node(abc) {};
	\draw (0.75, 0.75) node(abcd) {};
	\draw (abcd) -- (abc);
	\draw (abcd) -- (d);
	\draw (abc) -- (ab);
	\draw (abc) -- (c);
	\draw (ab) -- (a);
	\draw (ab) -- (b);
\end{scope}
\draw[<->] (-2.440420, 5.481273) -- (-1.247470, 5.868886);
	\draw (-3.750000, 6.245191) node[fill={none}, minimum size=5mm] {12};
\draw[<->] (3.897114, 2.250000) -- (-3.897114, -2.250000);
\draw[<->] (-3.897114, 2.250000) -- (3.897114, -2.250000);
\draw[<->] (0.000000, 4.500000) -- (0.000000, -4.500000);
\end{tikzpicture}%
  \endpgfgraphicnamed%

  \caption{The 12-cycle describing the walk in tree space
    corresponding to moving along the sequence $T'_0, T'_1, \ldots,
    T'_k$.}
  \label{fig:12cycle}
\end{figure}
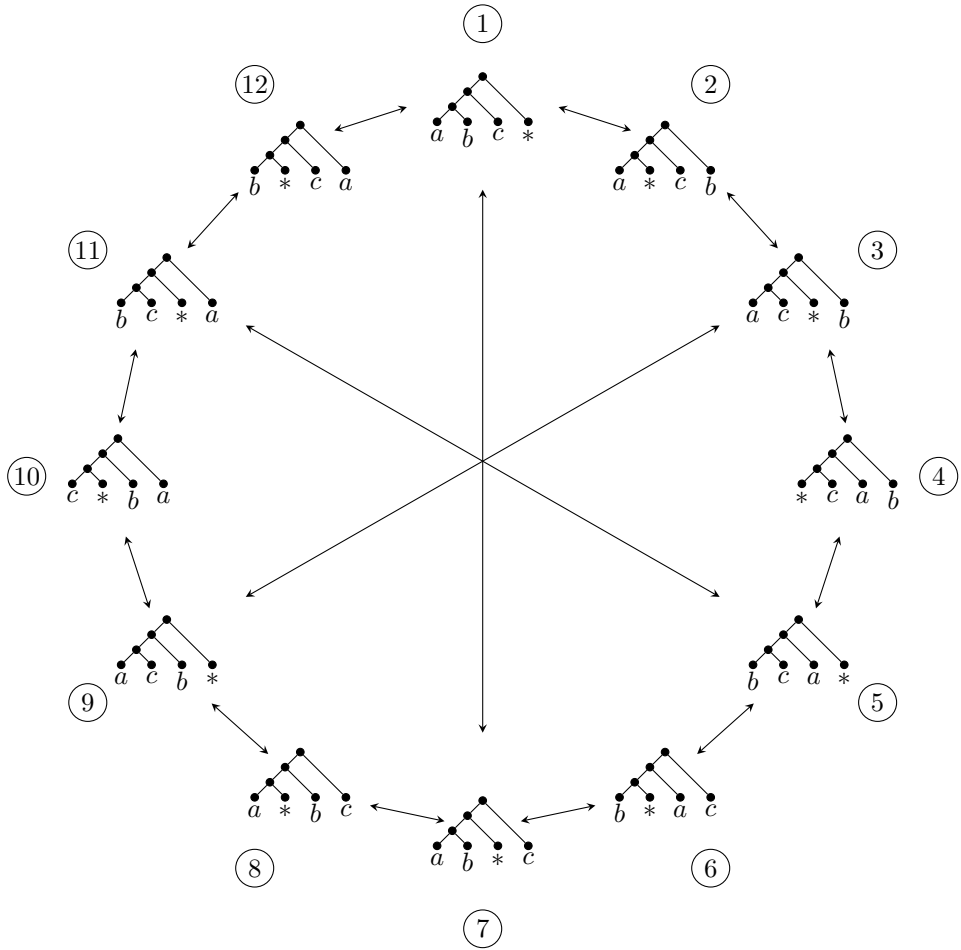

{$\boldsymbol{\mathbb{P}(\xi | J = 0)}$:} When $J = 0$,
Lemma~\ref{lemma:match_mismatch}.\ref{lemma:no_A_joining} tells us
that we need to look at the $A$--moving transfers, and
Lemma~\ref{lemma:T_prime_topology} tells us that $T'_k$, as described
in the preamble of the two lemmas, will induce the same topology on
$A$ as $T$. The process of $A$--moving transfers between time $t = 0$
and $t = t_2$ is a Poisson process in which the rate at which any
given $x \in A$ is moved is $\frac{1}{3}\lambda$, since each of the
three $A$--lineages can be moved to only one ($*$) out of three other
lineages (otherwise it would be an $A$--joining transfer). Note that
this process is independent of $J$ as the source point of an
$A$--joining transfer will always have an element of $A$ as a
descendant, whereas the source point of an $A$--moving transfer will
not. The walk in tree space, corresponding to moving along the
sequence $T'_0, T'_1, \ldots, T'_k$, as this process proceeds is
described by the Markov chain illustrated in Fig.~\ref{fig:12cycle}
with rate $\frac{1}{3}\lambda$ of moving from any state to each of its
neighbors.

Now let $Z_t : t \ge 0$ be a continuous-time symmetric random walk on
the $12$-cycle illustrated in Fig.~\ref{fig:12cycle}, where the
instantaneous rate of moving from one node to one of its neighbors is
$\frac{1}{3}\lambda$. As $T$ has topology $(ab;c;*)$ the random walk's
initial state is state $1$. The 7-state model, treated in the previous
section, is obtained from the model in Fig.~\ref{fig:12cycle} by
grouping state $2$ and $12$, $3$ and $11$, $4$ and $10$, $5$ and $9$,
and $6$ and $8$. So, accordingly, let $p_r(t)$ for $r=0,1, \ldots, 6$
be the probability that, after running this process for time $t$,
$Z_t$ is at a state that can be reached in $r$ steps from state $1$,
taking no diagonal edges, and let $\mathbf{p}(t) = (p_0(t), p_1(t),
p_2(t), p_3(t), p_4(t), p_5(t), p_6(t))$. Then $\mathbf{p}(0) =
(1,0,0,0,0,0,0)$ and $\mathbf{p}(t)$ behaves as described in
Lemma~\ref{lemma:matrix_representation1} after rescaling time by a
factor $\frac{1}{3}\lambda$.

Let $\tau_i$ be the state of the random walk on the 12-cycle at time
$t_2$. Lemma~\ref{lemma:T_prime_topology} then ensures, that if
$\underline{\sigma}'$ is the sequence of $A$-moving transfers between
$t = 0$ and $t = t_2$, then $T[\underline{\sigma}']$ resolves $a, b$
and $c$ in the same way as $\tau_i$ does. At time $t = t_2$ the
random walk on the $12$-cycle is in one of the following states:
\begin{itemize}
\item{
    $1$ in which case $T[\sigma]|A = ab|c$ with probability $1$ regardless of any LGT events after $t_2$.
  }
\item{
    $2$ or $12$ in which case 
    \begin{itemize}
    \item{$T[\sigma]|A = ab|c$ with probability $\frac{1}{3}$ if there is at least one transfer event between $t_2$ and $t_3$, and}
    \item{$T[\sigma]|A = ac|b$ and $T[\sigma]|A = bc|a$ both have probability
        \begin{itemize}
        \item{$\frac{1}{2}$ if there is no LGT events between $t_2$ and $t_3$, and}
        \item{$\frac{1}{3}$ if there is at least one transfer between $t_2$ and $t_3$.}
        \end{itemize}
      }
    \end{itemize}
  }
\item{
    $3$ or $11$ in which case $T[\sigma]|A = ac|b$ and $T[\sigma]|A = bc|a$ both have probability $\frac{1}{2}$ regardless of any LGT events after $t_2$.
  }
\item{
    $4$ or $10$ in which case 
    \begin{itemize}
    \item{$T[\sigma]|A = ab|c$ with probability $\frac{1}{3}$ if there is at least one transfer event between $t_2$ and $t_3$, and}
    \item{$T[\sigma]|A = ac|b$ and $T[\sigma]|A = bc|a$ both have probability
        \begin{itemize}
        \item{$\frac{1}{2}$ if there is no LGT events between $t_2$ and $t_3$, and}
        \item{$\frac{1}{3}$ if there is at least one transfer between $t_2$ and $t_3$.}
        \end{itemize}
      }
    \end{itemize}
  }
\item{
    $5$ or $9$ in which case $T[\sigma]|A = ac|b$ and $T[\sigma]|A = bc|a$ both have probability $\frac{1}{2}$ regardless of any LGT events after $t_2$.
  }
\item{
    $6$ or $8$ in which case 
    \begin{itemize}
    \item{$T[\sigma]|A = ab|c$ with probability
        \begin{itemize}
        \item{$1$ if there is no LGT events between
            $t_2$ and $t_3$, or}
        \item{$\frac{1}{3}$ if there is at least one
            transfer event between $t_2$ and $t_3$, and}
        \end{itemize}
      }
    \item{$T[\sigma]|A = ac|b$ and $T[\sigma]|A = bc|a$ both have probability $\frac{1}{3}$ if there is at least one transfer between $t_2$ and $t_3$.}
    \end{itemize}
  }
\item{
    $7$ in which case $T[\sigma]|A = ab|c$ with probability $1$ regardless of any LGT events after $t_2$.
  }
\end{itemize}
Let $\mu = \frac{1}{3} \lambda t_2$ and $B = 3\lambda(t_3 -
t_2)$. Then the probability that there is no LGT event between $t_2$
and $t_3$ is $e^{-B}$, and the probability that there is at least one
LGT event in the same time span is thus $1 - e^{-B}$. Consequently we
get the following from combining the cases above:
\begin{equation}
  \begin{aligned}
    \mathbb{P}(\xi_{ab|c} | J = 0)   &= p_0(\mu) \cdot 1 \\
                                   &+ p_1(\mu) \cdot \frac{1}{3}(1 - e^{-B}) \\
                                   &+ p_2(\mu) \cdot 0 \\
                                   &+ p_3(\mu) \cdot \frac{1}{3}(1 - e^{-B}) \\
                                   &+ p_4(\mu) \cdot 0 \\
                                   &+ p_5(\mu) \cdot (1 \cdot e^{-B} + \frac{1}{3}(1 - e^{-B})) \\
                                   &+ p_6(\mu) \cdot 1\\
                                   &= p_0(\mu) + p_6(\mu) + p_5(\mu)e^{-B} + \frac{1}{3}(1 - e^{-B})(p_1(\mu) + p_3(\mu) + p_5(\mu))
  \end{aligned}
  \label{eq:P_xi_J_0}
\end{equation}
Similarly we get
\begin{equation}
  \begin{aligned}
    \mathbb{P}(\xi_{ac|b} | J = 0) = \mathbb{P}(\xi_{bc|a} | J = 0) &= \frac{1}{2}(p_2(\mu) + p_4(\mu)) + \frac{1}{2} e^{-B} (p_1(\mu) + p_3(\mu))\\
                                                                &+ \frac{1}{3}(1 - e^{-B}) (p_1(\mu) + p_3(\mu) + p_5(\mu)).
  \end{aligned}
\end{equation}
\noindent Now using Lemma~\ref{lemma:matrix_representation1}, we get
\begin{equation}
  \begin{aligned}
    \mathbb{P}(\xi_{ab|c} | J = 0) &= \frac{1}{6} e^{-4\mu} + \frac{1}{3}e^{-3\mu} + \frac{1}{3}e^{-\mu} + \frac{1}{6}\\
                        &+ (\frac{1}{4} e^{-5\mu} - \frac{1}{6} e^{-4\mu} - \frac{1}{6} e^{-3\mu} - \frac{1}{12} e^{-\mu} + \frac{1}{6})e^{-B}\\
                        &+ (\frac{1}{2} - \frac{1}{2} e^{-4\mu})\frac{1}{3}(1 - e^{-B})\\
                        &= \frac{1}{3} (1 + \frac{3}{4} e^{-5\mu} e^{-B} +  (1 - \frac{1}{2} e^{-B}) e^{-3\mu} +  (1 - \frac{1}{4} e^{-B}) e^{-\mu}),
  \end{aligned}
\end{equation}
\noindent and finally, using~\eqref{eq:P_xi} and~\eqref{eq:P_xi_J_gt_zero}, we arrive at
\begin{equation}
  \begin{aligned}
    \mathbb{P}(\xi_{ab|c}) &= \frac{1}{3}(1 + e^{-7\mu} (\frac{3}{4} e^{-B} e^{-4\mu} + (1 - \frac{1}{2} e^{-B}) e^{-2\mu} + (1 - \frac{1}{4} e^{-B}))).
  \end{aligned}
\end{equation}
From this it is easy to see that $\mathbb{P}(\xi) > \frac{1}{3}$ for all
positive values of $\mu$ and $B$, as $\frac{3}{4} e^{-B} e^{-4\mu} >
0$, $(1 - \frac{1}{2} e^{-B}) e^{-2\mu} > 0$ and $1 - \frac{1}{4}
e^{-B} > 0$. Hence since $\mathbb{P}(\xi_{ac|b}) = \mathbb{P}(\xi_{bc|a})$ we get $\mathbb{P}(\xi_{ab|c}) >
\frac{1}{3} > \mathbb{P}(\xi_{ac|b}) = \mathbb{P}(\xi_{bc|a})$.

\begin{figure}[h]
  \centering
  \includegraphics[scale=0.6]{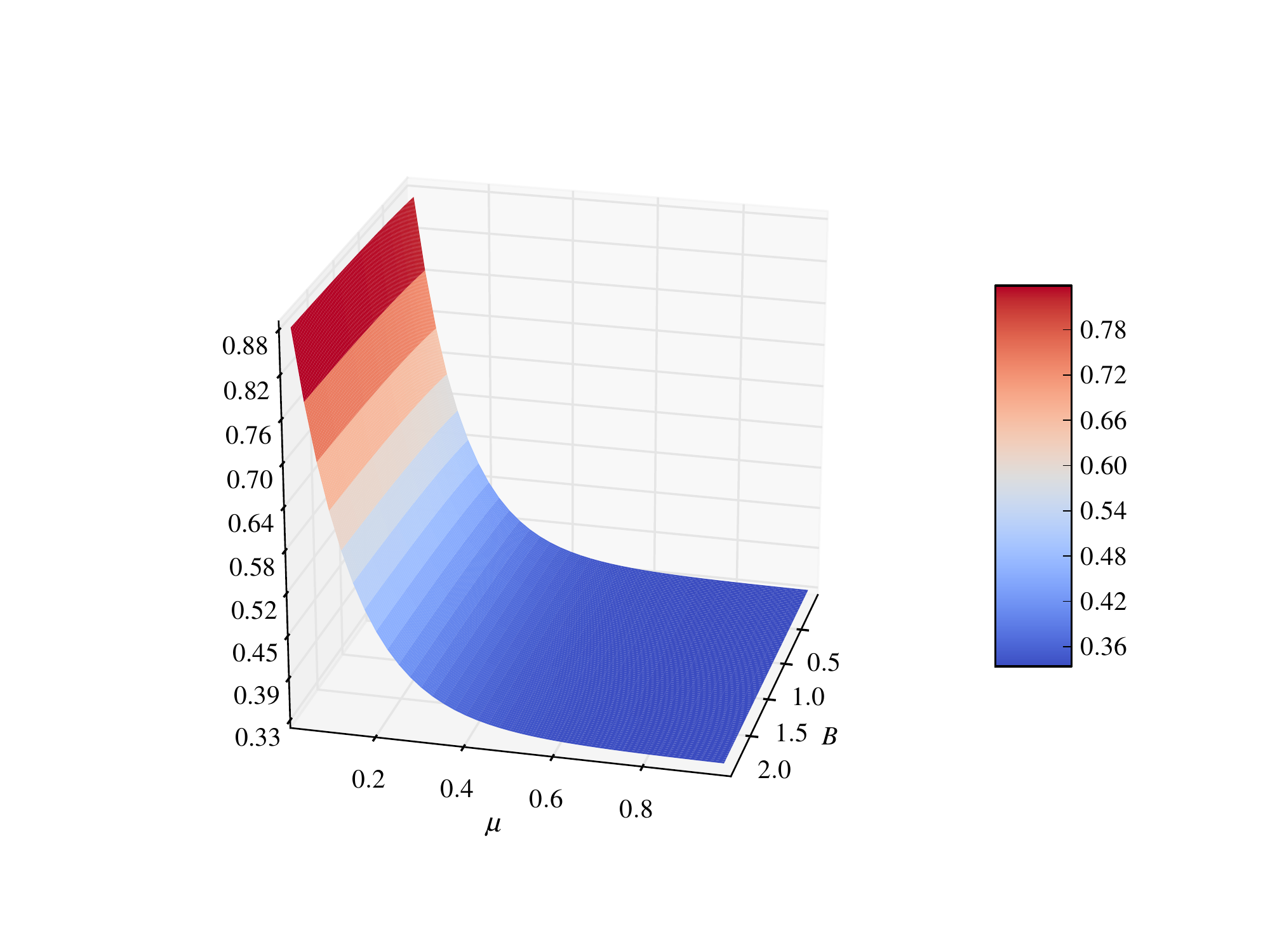}
  \caption{$\mathbb{P}(\xi_{ab|c})$ for four-taxon trees with the
    $(ab;c;*)$ topology as a function of $B = 3\lambda(t_3 - t_2)$ and
    $\mu = \frac{1}{3}\lambda t_2$.}
  \label{fig:abc_prob}
\end{figure}

A plot of $\mathbb{P}(\xi_{ab|c})$ as a function of $\mu$ and $B$ is shown in
Fig.~\ref{fig:abc_prob}.

\subsection{Proof of part \textit{(ii)}}
\label{sec:proof_ii}

The proof of the claim in part \textit{(ii)} of Theorem~\ref{theorem}
is completely analogous to the proof of part \textit{(i)} up until the
formulation of $\mathbb{P}(\xi_{a|bc})$ in~\eqref{eq:P_xi_J_0}. When
the original tree $T$ has topology $(ab;*;c)$, the random walk starts
in state $7$ (not state $1$ as it did before). Because of the
symmetries of the two Markov models, this means that $p_0(\mu)$ and
$p_6(\mu)$ swap places in~\eqref{eq:P_xi_J_0}, $p_1(\mu)$ and
$p_5(\mu)$ swap places, and $p_2(\mu)$ and $p_4(\mu)$ swap
places. Consequently we get
\begin{equation*}
  \mathbb{P}(\xi_{ab|c} | J = 0) = p_0(\mu) + p_6(\mu)  + p_1(\mu)e^{-B} + \frac{1}{3}(1 - e^{-B})(p_1(\mu) + p_3(\mu) + p_5(\mu))
  \label{eq:ii:xi_abc_J_0}
\end{equation*}
and $\mathbb{P}(\xi_{ac|b}) = \mathbb{P}(\xi_{bc|a})$. Now using
Lemma~\ref{lemma:matrix_representation1} we get
\begin{equation*}
  \begin{aligned}
    \mathbb{P}(\xi_{ab|c} | J = 0) = \frac{1}{3} ( 1 - \frac{3}{4}e^{-5\mu}e^{-B} + (1 - \frac{1}{2}e^{-B})e^{-3\mu} + (1 + \frac{5}{4}e^{-B})).
  \end{aligned}
\end{equation*}
And using~\eqref{eq:P_xi} and~\eqref{eq:P_xi_J_gt_zero} we arrive at
\begin{equation*}
  \mathbb{P}(\xi_{ab|c}) = \frac{1}{3} ( 1 - e^{-7\mu} (\frac{3}{4}e^{-4\mu}e^{-B} - (1 - \frac{1}{2}e^{-B})e^{-2\mu} - (1 + \frac{5}{4}e^{-B})  ) ).
\end{equation*}
We will now show that $\mathbb{P}(\xi_{ab|c}) > \frac{1}{3}$ for all $\mu, B >
0$. From the above we observe that $\mathbb{P}(\xi_{ab|c}) > \frac{1}{3}$ if
and only if
\begin{equation}
  \begin{aligned}
    &0 > \frac{3}{4}e^{-4\mu}e^{-B} - (1 - \frac{1}{2}e^{-B})e^{-2\mu} - (1 + \frac{5}{4}e^{-B}) \\
    ~&\Updownarrow\\
    &e^{-2\mu} + 1  > e^{-B} (\frac{3}{4}e^{-4\mu} + \frac{1}{2} e^{-2\mu} - \frac{5}{4}).
  \end{aligned}
 \label{eq:if_and_only_if}
\end{equation}
Now note that
\begin{equation*}
  \begin{array}{llll}
    \frac{3}{4}e^{-4\mu} + \frac{1}{2} e^{-2\mu} - \frac{5}{4} &\rightarrow &0\quad&\text{ for }\quad\mu \rightarrow 0\text{, and}\\
    & &\\
    \frac{3}{4}e^{-4\mu} + \frac{1}{2} e^{-2\mu} - \frac{5}{4} &\rightarrow &-\frac{5}{4}\quad&\text{ for }\quad\mu \rightarrow \infty.\\
  \end{array}
\end{equation*}
Thus $\frac{3}{4}e^{-4\mu} + \frac{1}{2} e^{-2\mu} - \frac{5}{4} < 0$
for all $\mu > 0$. This means that~\eqref{eq:if_and_only_if} is true
for all $\mu, B > 0$ since the left-hand side is always positive and
the right-hand side is always negative. We conclude that
$\mathbb{P}(\xi_{ab|c}) > \frac{1}{3}$ for all $\mu, B > 0$ and hence for all
values of $t_2$ and $t_3$.

\begin{figure}[h]
  \centering
  \includegraphics[scale=0.6]{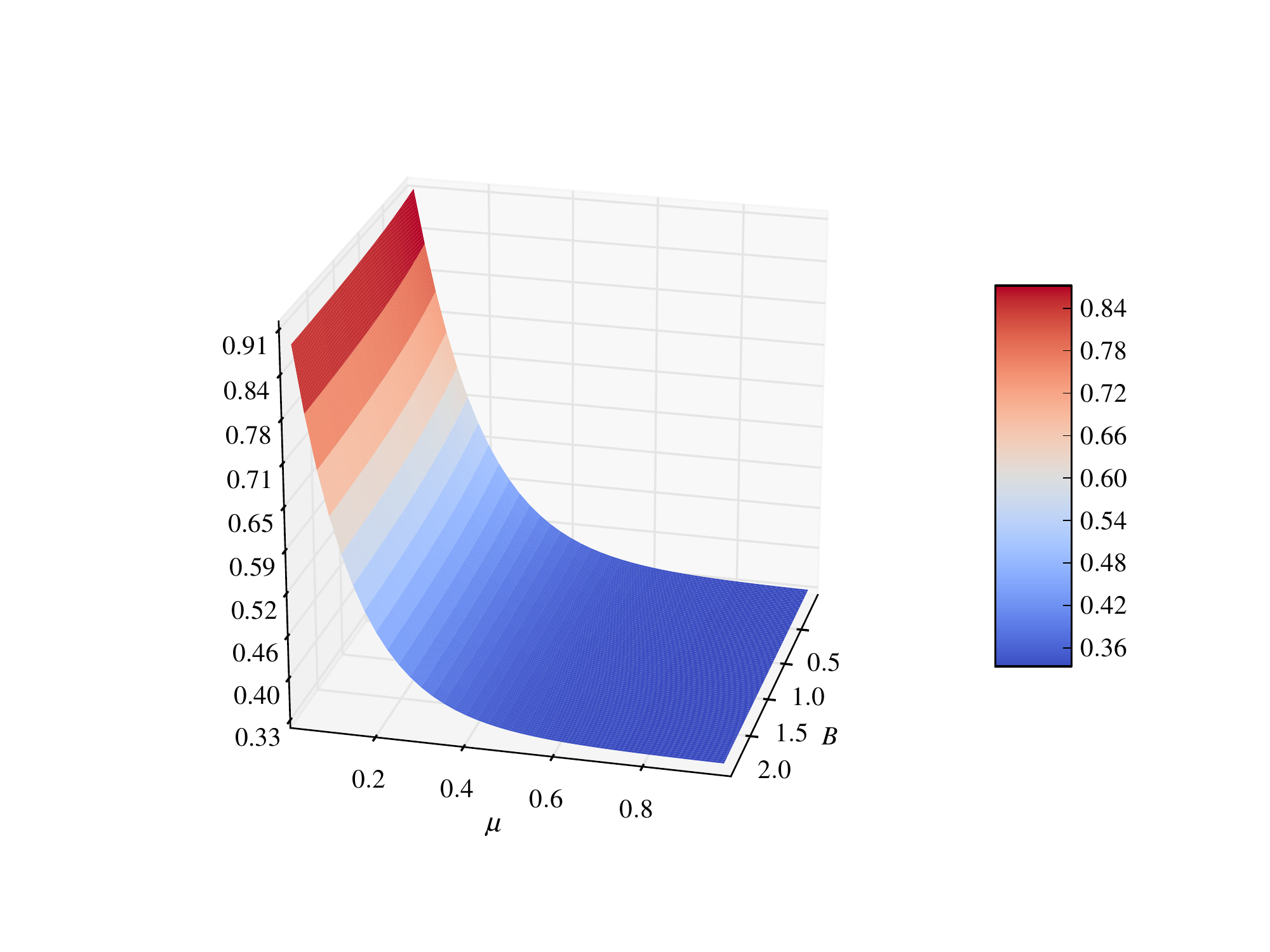}
  \caption{$\mathbb{P}(\xi_{ab|c})$ for four-taxon trees with the $(ab;*;c)$
    topology as a function of $B = 3\lambda(t_3 - t_2)$ and
    $\mu = \frac{1}{3}\lambda t_2$.}
  \label{fig:ab_c_prob}
\end{figure}

A plot of $\mathbb{P}(\xi_{ab|c})$ as a function of $\mu$ and $B$ is shown in
Fig.~\ref{fig:ab_c_prob}.

\subsection{Proof of part \textit{(iii)}}
\label{sec:proof_iii}
The proof of part \textit{(iii)} of Theorem~\ref{theorem} is
completely analogous to the proof of part \textit{(i)} and
\textit{(ii)} given in Section~\ref{sec:proof_i} and~\ref{sec:proof_i}
up until the computation of $\mathbb{P}(\xi | J=0)$.

As in the previous two sections let $Z_t : t \ge 0$ be a
continuous-time symmetric random walk on the 12-cycle illustrated in
Fig.~\ref{fig:12cycle}, where the instantaneous rate of moving from
one node to one of its neighbors is $\frac{1}{3}\lambda$. But this
time, as $T$ has topology $(a*;b;c)$ or $(b*;a;c)$, the random walk
starts in either state $6$ or $8$. Let $p_0(t), p_1(t), \ldots,
p_6(t)$ be defined as in Section~\ref{sec:proof_i} by
\begin{align*}
  p_0(t) & = \mathbb{P}(Z_t = 1) \\
  p_1(t) & = \mathbb{P}(Z_t = 2\text{ or }Z_t = 12) \\
  p_2(t) & = \mathbb{P}(Z_t = 3\text{ or }Z_t = 11) \\
  p_3(t) & = \mathbb{P}(Z_t = 4\text{ or }Z_t = 10) \\
  p_4(t) & = \mathbb{P}(Z_t = 5\text{ or }Z_t = 9) \\
  p_5(t) & = \mathbb{P}(Z_t = 6\text{ or }Z_t = 8) \\
  p_6(t) & = \mathbb{P}(Z_t = 7),
\end{align*}
and let $\mathbf{p}(t) = (p_0(t), p_1(t), p_2(t), p_3(t), p_4(t),
p_5(t), p_6(t))$. Then $\mathbf{p}(0) = (0,0,0,0,0,1,0)$ and
$\mathbf{p}(t)$ behaves as described in
Lemma~\ref{lemma:matrix_representation2} after rescaling time by a
factor $\frac{1}{3}\lambda$. As in Section~\ref{sec:proof_i} we get
\begin{equation*}
  \mathbb{P}(\xi_{ab|c} | J = 0) = p_0(\mu) + p_6(\mu) + \frac{1}{3}(1 - e^{-B})(p_1(\mu) + p_3(\mu) + p_5(\mu)) \ e^{-B}p_5(\mu)
\end{equation*}
and $\mathbb{P}(\xi_{ac|b} = \mathbb{P}(\xi_{bc|a})$. Using
Lemma~\ref{lemma:matrix_representation2} we now get
\begin{equation*}
  \mathbb{P}(\xi_{ab|c} | J = 0) = \frac{1}{3} (1 + \frac{3}{8}e^{-B}e^{-5\mu} - (\frac{1}{2} - \frac{1}{4}e^{-B})e^{-3\mu} + (\frac{1}{2} + \frac{11}{8}e^{-B})e^{-\mu}),
\end{equation*}
and we finally arrive at
\begin{equation*}
  \mathbb{P}(\xi_{ab|c}) = \frac{1}{3} (1 + e^{-7\mu}(\frac{3}{8}e^{-B}e^{-4\mu} - (\frac{1}{2} - \frac{1}{4}e^{-B})e^{-2\mu} + (\frac{1}{2} + \frac{11}{8}e^{-B}))).
\end{equation*}
To show that $\mathbb{P}(\xi_{ab|c}) > \frac{1}{3}$ for all $\mu, B > 0$ we
observe that $\mathbb{P}(\xi_{ab|c}) > \frac{1}{3}$ if and only if
\begin{equation}
  \begin{aligned}
    &0 < \frac{3}{8}e^{-B}e^{-5\mu} - (\frac{1}{2} - \frac{1}{4}e^{-B})e^{-3\mu} + (\frac{1}{2} + \frac{11}{8}e^{-B})e^{-\mu} \\
    ~&\Updownarrow \\
    &\frac{1}{2} (e^{-2\mu} - 1) < e^{-B} ( \frac{3}{8}e^{-4\mu} + \frac{1}{4}e^{-2\mu} + \frac{11}{8} ).
  \end{aligned}
  \label{eq:P_xi_gt}
\end{equation}
Note that
\begin{equation*}
  \begin{array}{llll}
    \frac{3}{8}e^{-4\mu} + \frac{1}{4}e^{-2\mu} + \frac{11}{8} &\rightarrow &2\quad&\text{ for }\quad\mu \rightarrow 0\text{, and}\\
    &&&\\
    \frac{3}{8}e^{-4\mu} + \frac{1}{4}e^{-2\mu} + \frac{11}{8} &\rightarrow &\frac{11}{8}\quad&\text{ for }\quad\mu \rightarrow \infty.\\
  \end{array}
\end{equation*}
Thus $\frac{3}{8}e^{-4\mu} + \frac{1}{4}e^{-2\mu} + \frac{11}{8} > 0$
for all $\mu > 0$. This means that~\eqref{eq:P_xi_gt} is true for all
$\mu, B > 0$, as the left-hand side is always negative, and the
right-hand side is always positive. We therefore conclude that
$\mathbb{P}(\xi_{ab|c}) > \frac{1}{3} > \mathbb{P}(\xi_{ac|b}) = \mathbb{P}(\xi_{bc|a})$ for all
values of $t_2$ and $t_3$.

\begin{figure}[h]
  \centering
  \includegraphics[scale=0.6]{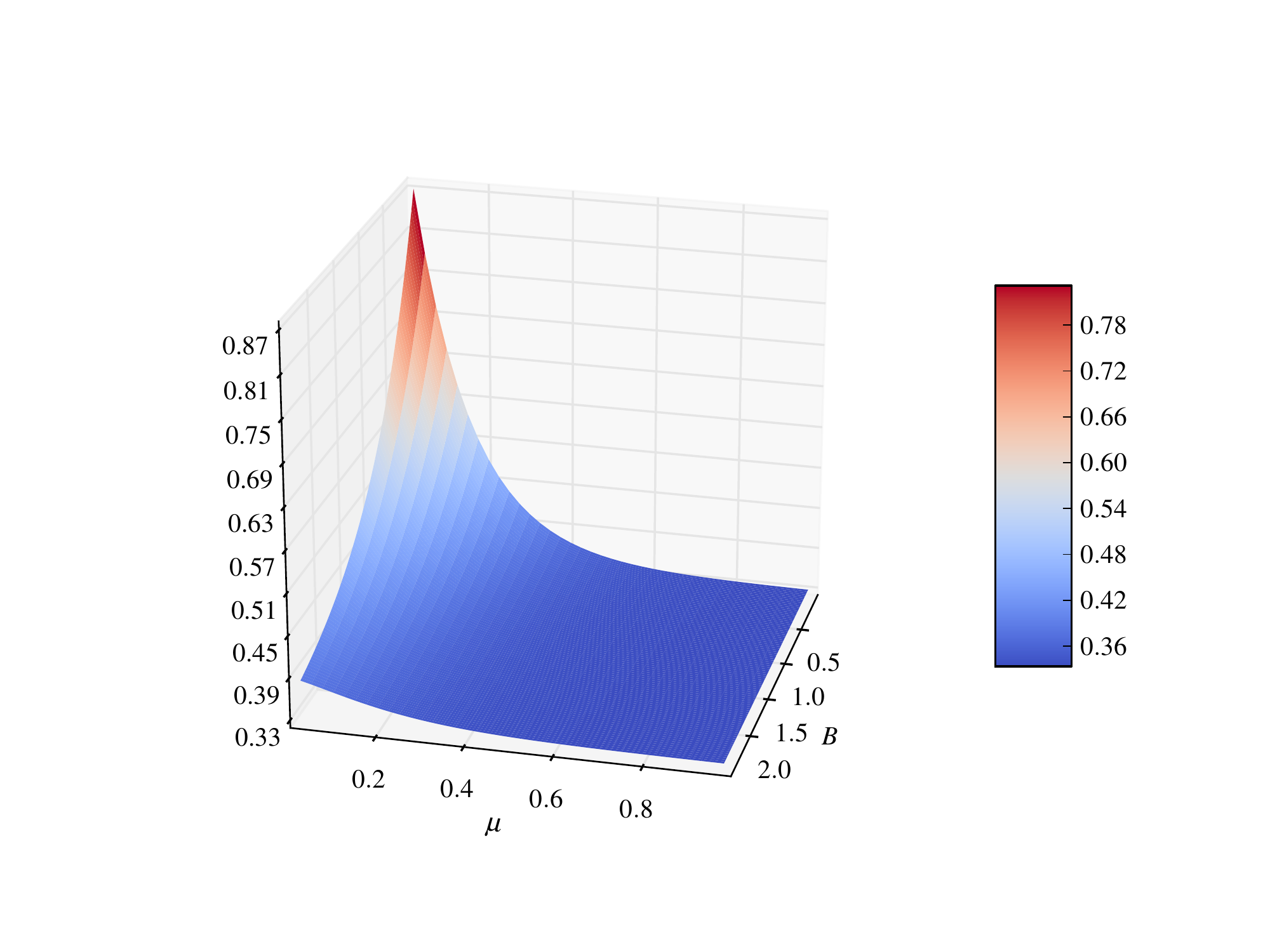}
  \caption{$\mathbb{P}(\xi_{ab|c})$ for four-taxon trees with the either the
    $(a*;b;c)$ or the $(b*;a;c)$ topology as a function of $B =
    3\lambda(t_3 - t_2)$ and $\mu = \frac{1}{3}\lambda
    t_2$.}
  \label{fig:a_bc_prob}
\end{figure}

A plot of $\mathbb{P}(\xi_{ab|c})$ as a function of $\mu$ and $B$ is shown in
Fig.~\ref{fig:a_bc_prob}.

\section{Limits}

It is interesting and reaffirming to study the limits of the
probabilities stated in Theorem~\ref{theorem} when $t_2$, $t_3$ or
$t_3-t_2$ approaches $0$. When $t_3$ approaches $0$ we leave no time
for any transfer events to occur before the first three taxa have
coalesced. We therefore expect to see that the triplet topology from
the species tree is preserved. And indeed $\mathbb{P}(\xi_{ab|c})
\rightarrow 1$ and $\mathbb{P}(\xi_{ac|b}) = \mathbb{P}(\xi_{bc|a})
\rightarrow 0$ as $t_3 \rightarrow 0$ in all three cases of
Theorem~\ref{theorem} (topology $(ab;c;*)$, $(ab;*;c)$ and $(a*;b;c)$
or $(b*;a;c)$).

When $t_2$ approaches $0$ we leave no time for any transfer events,
splitting up the two most closely related taxa, to occur, since such
events would have to happen before time $t_2$. Thus we expect that the
grouping of these two is preserved from the species tree topology. We
can recognize this behaviour in the limits for the topologies
$(ab;c;*)$ and $(ab;*;c)$, where $\mathbb{P}(\xi_{ab|c}) \rightarrow
1$ and $\mathbb{P}(\xi_{ac|b}) = \mathbb{P}(\xi_{bc|a}) \rightarrow 0$
as $t_2$ approaches $0$. Taxa $a$ and $b$ are here invariably grouped
together, and the matching topology is preserved regardless of any
transfer events after time $t_2$. If the species tree has topology
$(a*;b;c)$ or $(b*;a;c)$ then the species tree's triplet topology is
preserved with probability $1$ if no transfer events occur between
$t_2$ and $t_3$ and with probability $\frac{1}{3}$ if at least one
transfer event occurs between $t_2$ and $t_3$ (such an event would be
an $A$-joining event). Similarly a mismatching topology can only be
obtained if at least one transfer event occurs between $t_2$ and
$t_3$, in which case either of the two topologies is obtained with
probability $\frac{1}{3}$. Indeed we see from Theorem~\ref{theorem}
that $\mathbb{P}(\xi_{ab|c}) \rightarrow e^{-B} + \frac{1}{3}(1 -
e^{-B})$ and $\mathbb{P}(\xi_{ac|b}) = \mathbb{P}(\xi_{bc|a})
\rightarrow \frac{1}{3}(1 - e^{-B})$ as $t_2$ approaches $0$.

When $t_3-t_2$ approaches $0$ the triplet topology in a gene tree
entirely depends on the transfer events taking place before time
$t_2$. But since any kind of events can happen in this period of time,
this case is more complex than the two previous cases. The limits
obtained from the probabilities in Theorem~\ref{theorem} are as
follows:
\begin{itemize}
\item{
If $T$ has topology $(ab;c;*)$ then
\begin{align*}
  &\mathbb{P}(\xi_{ab|c}) \rightarrow \frac{1}{3}(1 + e^{-7\mu}(\frac{3}{4} + \frac{3}{4}e^{-4\mu} + \frac{1}{2}e^{-2\mu})) > \frac{1}{3}\; \forall \mu > 0 \text{, and}\\
  &\mathbb{P}(\xi_{ac|b}) = \mathbb{P}(\xi_{bc|a}) \rightarrow \frac{1}{3}(1 - \frac{e^{-7\mu}}{2}(\frac{3}{4} + \frac{3}{4}e^{-4\mu} + \frac{1}{2}e^{-2\mu}))  < \frac{1}{3}\; \forall \mu > 0.
\end{align*}
}
\item{
If $T$ has topology $(ab;*;c)$ then
\begin{align*}
  &\mathbb{P}(\xi_{ab|c}) \rightarrow \frac{1}{3}(1 + e^{-7\mu}(\frac{9}{4} - \frac{3}{4}e^{-4\mu} + \frac{1}{2}e^{-2\mu}))  > \frac{1}{3}\; \forall \mu > 0 \text{, and}\\
  &\mathbb{P}(\xi_{ac|b}) = \mathbb{P}(\xi_{bc|a}) \rightarrow \frac{1}{3}(1 - \frac{e^{-7\mu}}{2}(\frac{9}{4} - \frac{3}{4}e^{-4\mu} + \frac{1}{2}e^{-2\mu}))  < \frac{1}{3} \forall \; \mu > 0.
\end{align*}
}
\item{
If $T$ has topology $(a*;b;c)$ or $(b*;a;c)$ then
\begin{align*}
  &\mathbb{P}(\xi_{ab|c}) \rightarrow \frac{1}{3}(1 + e^{-7\mu}(\frac{15}{8} + \frac{3}{8}e^{-4\mu} - \frac{1}{4}e^{-2\mu}))  > \frac{1}{3} \; \forall \mu > 0 \text{, and}\\
  &\mathbb{P}(\xi_{ac|b}) = \mathbb{P}(\xi_{bc|a}) \rightarrow \frac{1}{3}(1 - \frac{e^{-7\mu}}{2}(\frac{15}{8} + \frac{3}{8}e^{-4\mu} - \frac{1}{4}e^{-2\mu}))  < \frac{1}{3} \; \forall \mu > 0.
\end{align*}
}
\end{itemize}
It is interesting to note that the probability of the triplet topology
matching the species tree always approaches a value strictly greater
than $\frac{1}{3}$ as $t_3-t_2$ approaches $0$. This is in contrast to
more familiar stochastic processes in phylogenetics -- such as lineage
sorting and site substitution models -- where shrinking an interior
branch length to zero results in a convergence to $\frac{1}{3},
\frac{1}{3}, \frac{1}{3}$ in support for the three resolutions of the
resulting trifurcation.

\nocite{*}
\bibliography{note} \bibliographystyle{plain}

\end{document}